\documentclass[superscriptaddress,twocolumn,showpacs,
amsmath,amssymb,prl,floatfix]{revtex4}

\usepackage{graphicx}
\usepackage{dcolumn} 
\usepackage{bm} 

\newcommand{\bnd}{$b_{nd}= (6.665 \pm 0.004)$~fm}
\newcommand{\bndowa}{$b_{nd}= (6.6730 \pm 0.0045)$~fm}
\newcommand{\bndnwa}{$b_{nd}= (6.669 \pm 0.003)$~fm}

\begin{document}

    \title{Precision neutron interferometric measurement of the $nd$
    coherent neutron \\ scattering length and consequences for models
    of three-nucleon forces}

\author{T.\,C. Black}
\affiliation{University of North Carolina at Wilmington, Wilmington,
NC 28403-3297, USA} 
\author{P.\,R. Huffman}
\affiliation{National Institute of Standards and Technology,
Gaithersburg, MD 20899-8461, USA} 
\author{D.\,L. Jacobson}
\affiliation{National Institute of Standards and Technology,
Gaithersburg, MD 20899-8461, USA} 
\author{W.\,M. Snow} 
\affiliation{Indiana University/IUCF, Bloomington, IN 47408, USA} 
\author{K.~Schoen} 
\affiliation{University of Missouri-Columbia, Columbia, MO 65211, USA} 
\author{M.~Arif}
\affiliation{National Institute of Standards and Technology,
Gaithersburg, MD 20899-8461, USA} 
\author{H.~Kaiser} 
\affiliation{University of Missouri-Columbia, Columbia, MO 65211, USA} 
\author{S.\,K. Lamoreaux}
\affiliation{Los Alamos National Laboratory, Los Alamos, NM 87545,
USA} 
\author{S.\,A. Werner}
\affiliation{University of Missouri-Columbia, Columbia, MO 65211, USA}
\affiliation{National Institute of Standards and Technology,
Gaithersburg, MD 20899-8461, USA}

\date{\today}

\begin{abstract}
We have performed the first high precision measurement 
of the coherent neutron scattering length of deuterium in a pure sample using neutron interferometry.  We find \bnd\ in agreement with the world average of
previous measurements using different techniques, \bndowa.  We compare
the new world average for the $nd$ coherent scattering length \bndnwa\
to calculations of the doublet and quartet scattering lengths from
several modern nucleon-nucleon potential models with three-nucleon
force (3NF) additions and show that almost all theories are in serious
disagreement with experiment.  This comparison is a more stringent
test of the models than past comparisons with the less precisely-determined doublet scattering length of $^{2}a_{nd}= (0.65 \pm 0.04)$~fm.  
\end{abstract}

\pacs{03.75.Dg, 21.45.+v, 25.10.+s, 25.40.Dn, 21.30.-x}

\maketitle

The nature of three-nucleon forces (3NF) has been an especially active
area of study in nuclear physics over the last decade and a
half\cite{Can00}.  Nonetheless, the current models are incomplete and
have a tendency to resolve discrepancies in some observables at
the expense of exacerbating discrepancies in others\cite{Kur02,Glo96}. 
In this Letter, we present a new measurement of the coherent $nd$
scattering length, $b_{nd}$, which can be used as part of a stringent
program of tests of NN and 3NF models.  We show that almost all modern
nucleon-nucleon (NN) potentials, including those with parametric 3NF
adjusted to replicate the triton binding energy, disagree with the new
world average value of $b_{nd}$.

The present family of modern NN potentials---AV18\cite{Wir95},
CD-Bonn\cite{Mac01}, and the various Nijmegen
potentials\cite{Sto94}---fit the extensive database of NN bound state
properties and scattering observables to within the precision of the
data.  More recent potential models based on chiral perturbation
theory predict NN properties and observables with comparable
accuracy\cite{Ent02,Epe01}.  There is growing confidence that the NN
interaction is understood well enough that deviations from NN force
model predictions in 3N observables for which relativistic corrections
are negligible can confidently be interpreted as 3NF effects.

The simplest systems that exhibit 3NF effects are the bound states of
$^{3}$H and $^{3}$He and the scattering states of $nd$ and $pd$.  Since
$^{3}$H and $nd$ are free from long-range electromagnetic interactions
that complicate both theory and experiment, they are the systems of
choice for precision tests at low energy.  The computational tools 
presently available to analyze these systems are believed to be
excellent.  

The 3N interactions employed in realistic calculations have evolved to
reconcile a growing number of discrepancies between theoretical and
experimental observables\cite{Pie01}; the nucleon binding energies,
the nucleon-deuteron ($Nd$) doublet scattering lengths, the $Nd$
vector analyzing power $A_y$ at low energies\cite{Tor98}, and $Nd$
spin observables above the deuteron breakup threshold\cite{Ste99}. 
Although the trinucleon binding energy and the $Nd$ doublet scattering
lengths $a_{Nd}$ have been observed to be computationally correlated, as shown in the well-known Phillips lines, they are not equivalent.  The doublet scattering length has special importance for effective field theory (EFT) calculations of 3N observables.  A momentum cut-off parameter required in the renormalized theory can be adjusted to match the experimental value of $^2a_{Nd}$\cite{Dil71}, which then forms the fundamental expansion parameter for the system in this channel\cite{Bed99}.

The $nd$ bound state coherent scattering length $b$ is the forward scattering amplitude due to the strong interaction from an unpolarized ensemble in the limit of zero neutron energy.  It is simply related to the free nuclear doublet and quartet scattering lengths $^{2S+1}a_{nd}$\cite{Samsbook}
\begin{equation}
\label{eq:comp}
b_{nd} = \left(\frac{m_n + m_d}{m_d}\right)\left[ \left(\tfrac{1}{3}\right)~^{2}a_{nd} + \left(\tfrac{2}{3}\right)~^{4}a_{nd}\right], 
\end{equation} 
where $m_{n}$ and $m_{d}$ are the neutron and deuteron masses.  

Using the Neutron Interferometer and Optics Facility (NIOF) at
NIST\cite{Ari94}, we have measured $b_{nd}$ using a pure D$_{2}$ gas
sample.  Since the quartet scattering length can be accurately
calculated from NN potentials and is largely determined by deuteron
properties and therefore insensitive to 3NF effects, we believe that 
comparison to the coherent $nd$ scattering length is a more stringent test of NN+3NF models than comparison to the doublet scattering length.  Moreover, since the effective range function in the quartet spin channel is a smooth
function of energy, the quartet scattering length can be accurately
extracted from an energy dependent phase shift analysis\cite{Bla99},
so that both theoretically and experimentally, knowledge of the bound
coherent scattering length is equivalent to knowledge of the doublet
scattering length.

In neutron interferometry \cite{Samsbook}, the phase shift of the
neutron beam due to the optical potential of a material is measured. 
To high accuracy, the phase shift due to the sample is 
\begin{equation}
    \label{eq:phase}
    \Delta \phi = (n_{r} - 1) k D_{\mathit{eff}} 
                = - \sum_{l} \lambda N_{l} b_{l} D_{\mathit{eff}},
\end{equation}
where $n_{r}$ is the real part of the index of refraction, $(n_{r}- 1)
\approx 10^{-5}$, $k$ is the wavevector, and the sum is taken over
the elemental species.  To measure the bound coherent scattering
length $b$ to $0.05$~\% absolute accuracy, measurements of the neutron
optical phase shift $\Delta \phi$, the atom density $N_{l}$, the
sample thickness $D_{\mathit{eff}}$, the neutron wavelength $\lambda$,
and the gas purity, at the $0.02$~\% level are required.
 
\begin{figure}
    \begin{center}
    \includegraphics{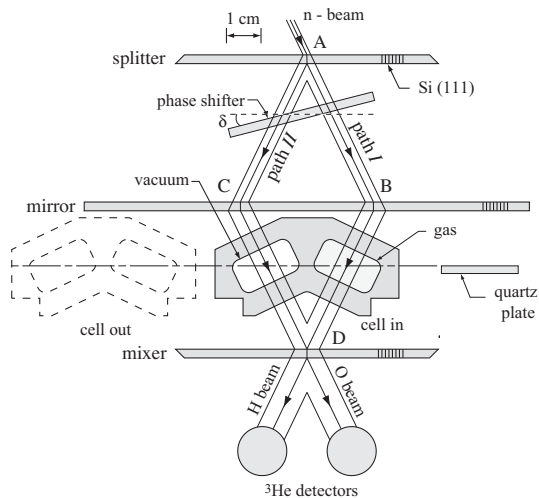}
    \end{center}
	\caption{A schematic view of the Si perfect crystal neutron
	interferometer with gas cell and quartz alignment flag. 
	Individual components are discussed in the text.}
    \label{fig:Interferometer}
\end{figure}

Measurements were performed using a perfect crystal silicon neutron
interferometer with high phase contrast (80~\%) and long-term phase
stability ($\approx$ 5$^{\circ}$ per day)\cite{Ari94}.  The single
crystal interferometer is schematically illustrated in
Fig.~\ref{fig:Interferometer}.
A monochromatic cold neutron beam ($E = 11.1$~meV, $\lambda =
0.271$~nm, $\Delta \lambda / \lambda \leq 0.5~\%$) is coherently
divided near point A by Bragg diffraction into two beams that travel
along paths $I$ and $II$.  These beams are again split near points B
and C and are coherently recombined to interfere near point D.

The phase shift, $\Delta \phi$, is measured by a secondary sampling
method in which a phase shifter is positioned across both beams.  The
intensities of the O and H beams as a function of the phase flag
angle, $\delta$, can be described by the following relations:
\begin{eqnarray}
    \label{eq:IO}
    I_{O}(\delta) & = & A_{O} + B \cos\left( C \xi(\delta)
             + \Delta\phi_{0}\right), \\
    I_{H}(\delta) & = & A_{H} + B \cos\left( C \xi(\delta)
             + \Delta\phi_{0} + \pi \right). \nonumber
\end{eqnarray}
The function $\xi(\delta)$ is the path length difference of the two
beams (I and II) traversing the phase flag.  The parameters $A_{O}$,
$A_{H}$, $B$, $C$, and $\Delta\phi_{0}$ are extracted from fits to the
data.  The value of $\Delta\phi_{0}$ and its corresponding uncertainty
is used to determine the phase difference between the two interfering
beams.

In order to minimize the effect of the phase shift due to the aluminum cell, 
the cell walls were designed to extend across both beam paths, producing 
compensating phase shifts.  When the cell is perfectly aligned, the beams strike
both cell compartments perpendicular to their surfaces, minimizing
systematics.  The resulting cell phase shift was 
five times smaller than the gas phase shift, reducing
the relative contribution of the cell to the total phase shift uncertainty to $1.7 \times 10^{-4}$.  Alignment of the cell was performed 
by a phase shift measurement using a quartz plate mounted on the kinematic
mount of the cell, similar to the procedure used in Ref.~\cite{Iof98}.

Gas was then introduced into the chamber on path $I$, while the cell
on path $II$ was evacuated.  Interferograms with the cell first in the
beam and then translated out of the beam were collected to determine
the phase shift due to the gas and to the small difference in aluminum
thickness.  This phase difference is combined with measurements of
$N$, $D_{\mathit{eff}}$, $\lambda$, and the gas purity to extract
$b_{nd}$.

The atom density $N$ is determined using the ideal gas law with virial
coefficient corrections up to the third pressure virial coefficient. 
For deuterium, the virial coefficients have been measured with
sufficient accuracy to determine $N$ with a relative uncertainty of
0.001~\%\cite{Mic59,Dym80}.  The absolute temperature was measured
using two calibrated 100~$\Omega$ platinum thermometers which have an
absolute accuracy of $0.023~\%$ at 300~K. The pressure was measured
using a silicon pressure transducer capable of measuring the
absolute pressure to better than 0.01~\%. 

The gas cell was ($1.0016 \pm 0.0001$)~cm thick at an absolute
temperature of ($20.00 \pm 0.05)~^{\circ}$C. Dimensional changes in
the cell due to the $\approx 12$~bar pressure result in a change in
thickness at the center of the cell of less than 1~$\mu$m, which
amounts to a systematic effect on the thickness of less than 0.01~\%.

Both mass spectroscopy and Raman spectroscopy were employed to measure
impurities in the D$_{2}$ gas sample, with the primary contaminants
being HD and D$_{2}$O. The mole fractions of D, H, and O were measured
to be $x_{D} = 0.99840 \pm 0.00017$, $x_{H} = 0.001500 \pm 0.000065$,
and $x_{O} = 0.000050 \pm 0.000016$.  The expression for $b_{nd}$,
corrected for impurities is, $b_{nd} = [b_{gas} - b_{H} x_{H} - b_{O}
x_{O}] / x_{D}$, where $b_{H} = (-3.746\pm0.0020)$~fm and $b_{O} =
(5.805\pm0.004)$~fm\cite{Koe91}.  Corrections due to the molecular state of the gas are not relevant within the reported accuracy of the measurement.  

The neutron wavelength was measured using an analyzer crystal placed
in the H-beam as described in Ref.~\cite{Lit97}.  Rotating this
crystal through both the symmetric and anti-symmetric Bragg
reflections allows the absolute Bragg angle, $\theta_{B}$, to be
determined to high accuracy, yielding $\lambda = (0.271266 \pm
0.000012)$~nm.  The stability of the wavelength over time was shown in
a separate test to be better than 0.001~\%.

\begin{figure}
    \includegraphics{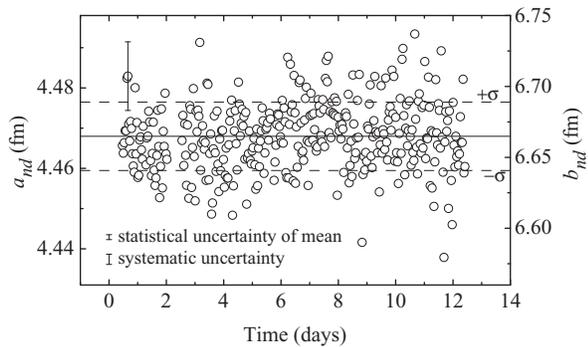}
    \caption{Values for the coherent scattering length on a run-to-run
    basis.}
    \label{fig:bd}
\end{figure}

\begin{figure}
    \includegraphics{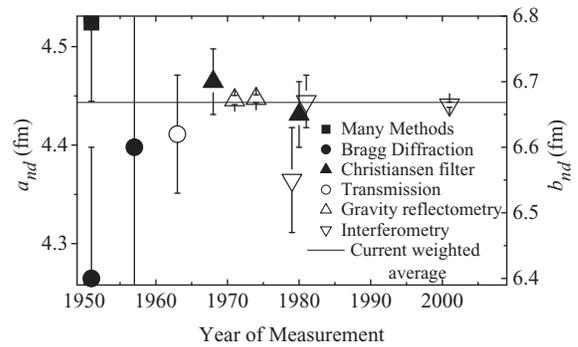}
    \caption{Bound coherent neutron scattering length for $nd$ 
	along with reported uncertainties.  Our result for D$_{2}$
	value is consistent with the previous and the current world
	average.}
    \label{fig:ScatteringLengthHistory}
\end{figure}

The value of $b_{nd}$ was extracted for each 42 min data set on a
run-by-run basis (Fig.~\ref{fig:bd}).  These values were then averaged
to obtain the coherent neutron scattering length \bnd.
To compare this value with the previous world average value of
$b_{nd}$, we consulted the existing compilations of previous
measurements\cite{Koe91,Rau00} and excluded all measurements which
were not published in a refereed journal and all measurements which
were later retracted.  The remaining values were combined into an
average with results weighted by the inverse square of their 
uncertainties in the usual manner\cite{Hag02}.  The values
are presented in Fig.~\ref{fig:ScatteringLengthHistory}.  The result
of this evaluation is \bndowa\ with a reduced $\chi ^{2}=0.6$.  Our
value, \bnd, is consistent with the previous world average and of
comparable precision.  The new world average value for the bound
nuclear scattering length of deuterium is \bndnwa.

For the forward scattering, which is the only component that
contributes to the phase shift seen by the interferometer, the
scattering amplitudes from both the Mott-Schwinger and
neutron-electron interactions are zero\cite{Samsbook}.  The size of
the scattering due to the electric polarizability of the neutron is
less than $-0.000017$~fm\cite{Sch91}.  This measurement is thus
sensitive only to the same nuclear interactions that are calculated in  
theoretical models of 3N scattering.

\begin{table}[t]
    \caption{Theoretical calculations of the $nd$ scattering lengths. 
    The rightmost column is the coherent scattering length.  The 3NF
    parameters in the boldfaced potential models have been adjusted to
    reproduce the triton binding energy.}
    \begin{tabular}{llrrr}
	{\bf Ref.} & \multicolumn{1}{c}{{\bf Potential model}} &
	\multicolumn{1}{c}{$^{2}a_{nd}$ [fm]} &
	\multicolumn{1}{c}{$^{4}a_{nd}$ [fm]} &
	\multicolumn{1}{c}{$a_{nd}$ [fm]} \\
	\hline \hline
	\cite{Fri83} & {\bf MT I--III} & 0.70 & 6.44 & 4.52 \\
	\hline
	\cite{Fri84} & AV14 & 1.35 & 6.38 & 4.70 \\
	\cline{2-5} \mbox{} & SSCC & 1.32 & 6.41 & 4.71 \\
	\hline \cite{Che91} & RSC & 1.52 & 6.302 & 4.71 \\
	\cline{2-5} \mbox{} & RSC+TM3NF & 0.393 & 6.308 & 4.336 \\
	\cline{2-5} \mbox{} & AV14 & 1.200 & 6.372 & 4.648 \\
	\cline{2-5} \mbox{} & AV14+BR3NF & 0.001 & 6.378 & 4.252 \\
	\cline{2-5} \mbox{} & {\bf RSC+TM3NF} & 0.657 & 6.304 & 4.422 \\
	\cline{2-5} \mbox{} & {\bf AV14+BR3NF} & 0.567 & 6.380 & 4.442 \\
	\hline \cite{Che89} & {\bf MTI--III} & 0.71 & 6.43 & 4.52 \\
	\hline \cite{Kie94} & {\bf MTI--III} & 0.702 & 6.442 & 4.529 \\
	\cline{2-5} \mbox{} & AV14 & 1.196 & 6.380 & 4.652 \\
	\hline \cite{Kie97b} & AV14 & 1.189 & 6.379 & 4.649 \\
	\cline{2-5} \mbox{} & {\bf AV14+TM3NF} & 0.5857 & 6.371 & 4.443 \\
	\hline \hline
    \end{tabular} 
    \label{tab1}
\end{table}

\begin{figure}[ht]
    \includegraphics{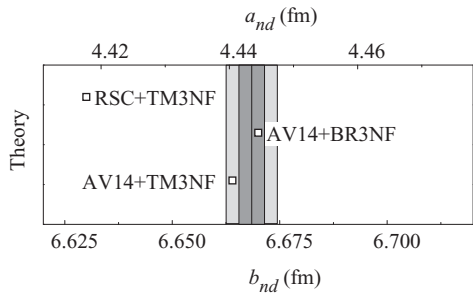}
	\caption{Theoretical calculations of the coherent scattering
	length compared with the new world average.  The central dark
	band is the 1$\sigma$ confidence band and the lighter band is
	the 2$\sigma$ confidence band.}
    \label{fig:TheoryCoherent}
\end{figure}

To illustrate the impact of comparing theoretical calculations to the
precision data on the coherent scattering length, a number of modern
calculations of the $nd$ scattering lengths are shown in
Table~\ref{tab1}.  The dependence of the theoretically calculated
$^2$S$_{1/2}$ scattering length on the inclusion of a 3NF is clearly
seen.  We note that, as expected, none of the theories which do not
incorporate a 3NF come close to matching the $nd$ coherent scattering
length.  The MTI-III model, which gives fair agreement without a 3NF, nonetheless 
fails to reproduce the NN partial wave phase shifts. Of the potential 
models that include a 3NF, only the AV14 potential with the Brazil 
3NF\cite{Che91} and the AV14 potential with the TM~3NF\cite{Kie97b} are in agreement with the data as shown in Fig.~\ref{fig:TheoryCoherent}. We wish to stress 
that the precision with which the coherent $nd$ scattering length is 
known can, in concert with the triton binding energy, set tight constraints on NN potential models as well as on 3BF models.  None of the potential models listed in 
Table~\ref{tab1}, for example, include charge independence breaking (CIB) effects.  
The authors of a recent paper\cite{Epe02} estimate that CIB effects increase the 
$nd$ doublet scattering length by 0.19~fm.  Arriving at a simultaneously correct NN 
and 3N potential model requires sufficient precision in all relevant fundamental low-energy parameters, especially the triton binding energy and the coherent $nd$ scattering length. 

We observe that none of the theories, with the exception of the AV14
potential with a 3N force, are in agreement with the precisely-known
world average coherent $nd$ scattering length of $a_{nd} = (4.443 \pm
0.002)$~fm.  The precision of the measurement of this critical
parameter is now such that we can distinguish between different
theoretical combinations of NN and 3NF models, even when the 3N forces
have been adjusted to replicate the triton binding energies.  Relativistic 
corrections to the triton binding energy, which were shown by the authors of 
\cite{Kam02} to weaken the binding by 0.29--0.43 MeV, further reduce the correlation between $b_{nd}$ and the triton binding energy, and provide another justification for using this pair of parameters to perform rigorous tests of 
NN and 3NF models.  Clearly the theoretical community now has an incentive to systematically investigate nuclear force models for these parameters with comparable accuracy, especially since the precision of interferometric measurements of neutron-nucleus scattering can be extended by at least an order of magnitude.  

We would like to thank P. Lawson and W. Dorko for assistance in this
work.  This work was supported by the U.S. Department of Commerce, the
National Science Foundation (NSF) Grant No.  PHY-9603559 at the
University of Missouri, and NSF Grant No.  PHY-9602872  at Indiana
University.


\begin{thebibliography}{00} 
\bibitem{Can00} L. Canton and W. Schadow, Phys. Rev. {\bf C62}, 
044005 (2000).
\bibitem{Kur02} J. Kuro\'s -\.Zo\l nierczuk, H. Wita\l a, J. Golak, H. Kamada, A. Nogga, R. Skibi\'nski, and W. Gl\"ockle, Phys. Rev. {\bf C66}, 024003 (2002).  
\bibitem{Glo96} W. Gl\"ockle, H. Wita\l a, D. H\"uber, H. Kamada, and J.
Golak, Phys. Rep. {\bf 274}, 107 (1996).
\bibitem{Wir95} R. B. Wiringa, V. G. J. Stoks, and R. Schiavilla,
Phys. Rev. {\bf C51}, 38 (1995).
\bibitem{Mac01} R. Machleidt, Phys. Rev. {\bf C63}, 024001 (2001).  
\bibitem{Sto94} V. G. J. Stoks, R. A. M. Klomp, C. P. F. Terheggen,
and J. J. deSwart, Phys. Rev. {\bf C49}, 2950 (1994). 
\bibitem{Ent02} D. R. Entem and R. Machleidt, Phys. Lett. {\bf B524}, 
(2002) 93. 
\bibitem{Epe01} E. Epelbaum, H. Kamada, A. Nogga, H. Wita\l a, W. 
Gl\"ockle,
and  Ulf-G. Mei\ss ner, Phys. Rev. Lett. {\bf 86}, 4787 (2001). 
\bibitem{Pie01} S. C. Pieper, V. R. Pandharipande, R. B. Wiringa, and 
J. Carlson, Phys. Rev. {\bf C64}, 014001 (2001).  
\bibitem{Tor98} W. Tornow and H. Wita\l a, Nucl. Phys. {\bf A637}, 280
(1998). 
\bibitem{Ste99} E. J. Stephenson, H. Wita\l a, W. Gl\"ockle, 
H. Kamada, and A. Nogga, Phys. Rev. {\bf C60}, 061001 (1999). 
\bibitem{Dil71} W. Dilg, L. Koester, W. Nistler, Phys. Lett. {\bf B36}, 208 (1971).   
\bibitem{Bed99} P. F. Bedaque, H. W. Hammer, and U. van Kolck, Phys. 
Rev. Lett. {\bf 82}, 463 (1999).  
\bibitem{Samsbook} H. Rauch, S. Werner, \textit{Neutron Interferometry:
Lessons in Experimental Quantum Mechanics}, Oxford University
Press, 2000.
\bibitem{Ari94}  M. Arif, D. E. Brown, G. L. Greene, R. Clothier, and
K. Littrell, Proc. SPIE Vol. 2264, edited by C. G. Gordon, pp.
21-26, 1994.
\bibitem{Bla99} T. C. Black, H. J. Karwowski, E. J. Ludwig, A.
Kievsky, M. Viviani, and S. Rosati, Phys. Lett. {\bf B471},103 (1999).  
\bibitem{Iof98} A. Ioffe, D. L. Jacobson, M. Arif, M. Vrana, S. A.
Werner, P. Fischer, G. L. Greene, and F. Mezei, Phys. Rev. {\bf
A58}, 1475 (1998).
\bibitem{Mic59} A. Michels, W. de Graaff, T. Wassenaar, J.M.H.
Levelt, and P. Louwerse, Physica Grav. {\bf 25}, 25 (1959).
\bibitem{Dym80} J. H. Dymond, \emph{The virial coefficients of pure gases
and mixtures: a critical compilation}, (New York, Oxford U. Press, 
1980).
\bibitem{Koe91} L. Koester, H. Rauch, and E. Seymann, At. 
and Nucl. Data Tab. {\bf 49}, 65 (1991).
\bibitem{Lit97} K. C. Littrell, Ph.D. Thesis, U. Missouri, 1997.
\bibitem{Rau00} H. Rauch and W. Waschkowski, in {\it Elementary Particles, 
Nuclei, and Atoms}, edited by H. Schopper, Landolt-Bornstein,
New Series, Group I, Vol. 16, Pt. A (Springer, Berlin, 2000), Chap. 6. 
\bibitem{Hag02} K. Hagiwara \textit{et al.}, {\it 2002 Review of 
Particle Physics}, available at http://pdg.lbl.gov/, Phys. Rev. {\bf D66}, 010001-1 (2002).
\bibitem{Sch91} J. Schmiedmayer, P. Riehs, J. A. Harvey, N. W. Hill,
Phys. Rev. Lett. {\bf 66}, 1015 (1991).
\bibitem{Fri83} J.L. Friar, B.F. Gibson, and G.L. Payne, Phys. Rev.
{\bf C28}, 983 (1983).
\bibitem{Fri84} J.L. Friar, B.F. Gibson, G.L. Payne, and C. R. Chen,
Phys. Rev. {\bf C30}, 1121 (1984). 
\bibitem{Che91} C. R. Chen, G.L. Payne, J.L. Friar, and B.F. Gibson,
Phys. Rev. {\bf C44}, 50 (1991).
\bibitem{Che89} C.R. Chen, G.L. Payne, J.L. Friar, and B.F. Gibson,
Phys. Rev. {\bf C39}, 1261 (1989).
\bibitem{Kie94} A. Kievsky, M. Viviani, and S. Rosati, Nucl. Phys.
{\bf A577}, 511 (1994).
\bibitem{Kie97b} A. Kievsky, Nucl. Phys. {\bf A624}, 125 (1997).
\bibitem{Epe02} E. Epelbaum, A. Nogga, W. 
Gl\"ockle, H. Kamada, Ulf-G. Mei\ss ner, and H. Wita\l a Phys. Rev. {\bf C66}, 064001 (2002). 
\bibitem{Kam02} H. Kamada, W. Gl\"ockle, J. Golak, and Ch. Elster Phys. Rev. 
{\bf C66}, 044010 (2002). 
\end{thebibliography}
\end{document}